\documentclass[aps,twocolumn,showpacs,prl]{revtex4}% Physical Review Letter

\RequirePackage{amsfonts}
\RequirePackage{amssymb}
\RequirePackage{amsmath}
\RequirePackage{graphicx}
\RequirePackage{makeidx}
\RequirePackage{ifthen}
\RequirePackage{color}

\newcommand{\be}{\begin{eqnarray}}          \newcommand{\ee}{\end{eqnarray}}
\newcommand{\ba}{\begin{align}}          \newcommand{\ea}{\end{align}}
\newcommand{\benum}{\begin{enumerate}}
\newcommand{\eenum}{\end{enumerate}}
\newcommand{\bitem}{\begin{itemize}}
\newcommand{\eitem}{\end{itemize}}

\definecolor{myblue}{rgb}{.1,.1,.7}
\definecolor{dcyan}{rgb}{.0,.6,.6}
\definecolor{dmagenta}{rgb}{0.6,0.0,0.6}
\definecolor{brown}{rgb}{0.6,0.2,0.}
\definecolor{darkblue}{rgb}{0,0,0.6}
\definecolor{darkred}{rgb}{0.75,0.0,0.0}
\definecolor{darkgreen}{rgb}{0.0,0.6,0.0}

\newcommand{\blue}{\color{blue}}

\newcommand{\usemarker}{Y}
\newcommand{\marker}[1]{
       \ifthenelse{\equal{\usemarker}{Y}}
                     {\mbox{}\marginpar{\tt #1}}{}
               }

\newcommand{\mk}[1]{
\ifthenelse{\equal{\usemarker}{Y}}
{\noindent\hskip -1truecm {\bf\blue$^{#1}$}}{}
}

\newcommand{\tmop}[1]{#1}
\begin{document}
\title{Implications of Charmless B Decays with Large Direct CP Violation}
\author{Yue-Liang Wu}
\affiliation{Institute of theoretical physics,Chinese Academy of Sciences\\
100080, Beijing, China}
%\email[Email]{ylwu@itp.ac.cn}
\author{Yu-Feng Zhou}
\affiliation{Dortmund University, 44221, Dortmund, Germany}
%\email[Email]{zhou@thoerie.physik.uni-muenchen.de}
%\date{\today}
\begin{abstract}
Based on the most recent data in charmless B decays including the
very recently reported large direct CP violations, it is shown
that the weak phase $\gamma$ can well be extracted without
two-fold ambiguity even only from two decay modes $\pi^+\pi^-$ and
$\pi^+K^-$, and its value is remarkably consistent with the global
standard model fit at a compatible accuracy. A fit to all the
$\pi\pi,\pi K$ data favor both  large electroweak penguin and
color-suppressed tree amplitude with large strong phases. It is
demonstrated that the inclusion of $\tmop{SU} ( 3 )$ symmetry
breaking effects of strong phases and the inelastic rescattering
effects can well improve the consistency of the data, while both
effects may not be sufficient to arrive at a small electroweak
penguin amplitude in the standard model. It is of interest to
notice that large or small electroweak penguin amplitude becomes a
testable prediction as they lead to significantly different
predictions for the direct CP violations for $\pi^0\pi^0$, $\pi^0
\bar{K}^0$ modes. Clearly, precise measurements on charmless $B$
decays will provide a window for probing new physics.
\end{abstract}
\preprint{DO-TH/0410}
%\keywords{}
\maketitle

The evidences of direct CP violation in $B$ decays have recently
been reported by the BaBar and Belle collaborations. The latest
averaged data give $a_{CP}(\pi^+\pi^-)=0.46\pm 0.13$ and
$a_{CP}(\pi^+ K^-)=-0.11\pm 0.02$\cite{newData}. Thus direct CP
violation has been established not only in the kaon system, but
also in B system. It has been shown that both direct CP violation
and $\Delta I = 1/2$ rule in kaon decays can be
understood in the standard model (SM)\cite{Wu:2000ki}. It is then
natural to test whether the observed direct CP violations and
decay rates in charmless B decays can be explained within the SM.
As the two experimental groups BaBar and Belle have reported more and more
accurate data for charmless B decays ($B\rightarrow \pi\pi, \pi
K$)\cite{HFaverage}, it then allows one to test the SM and to
explore possible indications for new physics, such as the
two-Higgs-doublet model with  spontaneous CP
violation\cite{S2HDM}, the supersymmetric models etc.
There have been several global analyzes which are based on either
model independent parameterizations such as quark flavor
diagrammatic decomposition \cite{SW,Buras,Rosner,Hocker}, isospin
decomposition\cite{Zhou:2000hg}, flavor SU(3) symmetry\cite{He},
or QCD inspired calculations such as QCD factorization
\cite{QCDFacFit,Hocker} and perturbative QCD approach\cite{pQCD}
as well as soft-collinear effective theory\cite{SCET}.

In this letter,  we are going to make a  step-by-step
fit for the charmless B decay modes based on
approximate SU(3) flavor symmetry and hierarchical structures of
diagrammatic amplitudes. Based on the most
recent data including the very recently
reported large direct CP violations, we arrive at the following
main observations:
i) the current data allow us to precisely extract the
weak phase $\gamma$ from only two modes $\pi^+\pi^-$ and $\pi^+ K^-$
without two-fold ambiguity. The resulting
numerical value of $\gamma$ is found to be
remarkably consistent with the global SM fit at a
compatible accuracy;
%Of particular, the extracted weak phase
%$\gamma$ is quite stable and not significantly modified by other
%possible effects;
% such as the SU(3) symmetry breaking and large
%electroweak penguin effects;
%
ii) A direct fit to all $\pi\pi$, $\pi K$ modes favors a large
electroweak penguin.  Furthermore, the large or small electroweak
penguin amplitude is found to be a testable prediction via
measuring direct CP violations in the decay modes $B\to \pi^0
\pi^0$ and $\pi^0 \bar{K}^0$ once more accurate data become
available.
%For small(large) electroweak penguin
%amplitudes, direct CP violations $a_{CP}(\pi^0\pi^0)$ and $a_{CP}(\pi^0K^0)$ are
%found to be large(small).
%This allows one to distinguish them in the recent
%future.
%
iii) all the amplitudes and strong phases in $B\rightarrow \pi\pi
, \pi K$ are extracted, which indicates large final state
interactions and non-factorizable QCD effects as the resulting
numerical results show an enhanced color-suppressed tree amplitude
and strong phase. It is shown that not only the large $\pi^0\pi$
branching ratios but also the large $\pi^0 K^0$ ones result in a
large color-suppressed tree amplitude with a large strong phase;
iv) it is the large $\pi^0 K$ branching ratio that mainly
responsible for a large electroweak penguin amplitude with a large
strong phase. In the case of a small electroweak penguin amplitude
fixed by the isospin relation in the SM, the resulting $\pi^0 K$
branching ratios are bellow the experimental data;
vi) SU(3) symmetry breaking of strong phases and $B\to DD$ rescattering effects
can well improve the consistency of the global fitting. However, it remains
necessary to have a large electroweak penguin amplitude with large strong
phases.

The diagrammatic decomposition approach is adopted to carry out a
global analysis. The advantage is that  in such an approach some decay
modes can form, in a good approximation, closed subsets, which
allows us to determine the relevant parameters without knowing the others.
Although the
number of data points decrease for each subset, the number of the
free parameters decrease as well. Of interest, the precision of
the determinations is not necessarily lower than that of the whole
global fit. Furthermore it may avoid the complicity and the
potential inconsistency in the current data when more decay modes
are involved in the whole global fit. The comparison between
different results from different subsets may provide us important
hints to understand those decays. In general, all the $B\to \pi
\pi$ decay modes can be written in terms of diagrammatic
amplitudes: tree ($\mathcal{T}$), color-suppressed tree
($\mathcal{C} )$, QCD penguin ($\mathcal{P}$), electro-weak
penguin ($\mathcal{P}_{\tmop{EW}}$), color suppressed electroweak
penguin ( $\mathcal{P}_{\tmop{EW}}^C$) etc. The corresponding
diagrams in $B\to \pi K$ are denoted by primed ones, such as
$\mathcal{T}'$, $\mathcal{P}'$, etc. Using  the CKM factors
$\lambda_q^{( s )} = V_{\tmop{qd} ( s )}^{\ast} V_{\tmop{qb}}$,
and the unitarity of the CKM matrix, the penguin type amplitude
can be decomposed as : $\mathcal{P}^{(')} = \lambda_u^{( s )}
\mathcal{P}_u + \lambda_c^{( s )} \mathcal{P}_c + \lambda^{( s
)}_t \mathcal{P}_t$. Defining $P_{} \equiv P_{\tmop{tc}}
=\mathcal{P}_t -\mathcal{P}_c$ , $P_{\tmop{tu}}\equiv
\mathcal{P}_t -\mathcal{P}_u$, $\hat{P}_{EW} = P_{EW} +
P_{EW}^C$ and factorize out the CKM factors, we arrive
at the following diagrammatic decomposition
\begin{align}
  \bar{A}_{\pi^+ \pi^-}
  & =  \lambda_u ( T - P_{\tmop{tu}} - \frac{2}{3}
  P_{\tmop{EW}, \tmop{tu}}^C ) - \lambda_c ( P + \frac{2}{3} P_{\tmop{EW}}^C )
  \nonumber\\
  \bar{A}_{\pi^- \pi^0}
  & =  - \frac{1}{\sqrt{2}} \left[ \lambda_u ( T + C -
  \hat{P}_{\tmop{EW}, \tmop{tu}} ) -
  \lambda_c  \hat{P}_{\tmop{EW}} \right]
  \nonumber\\
  \bar{A}_{\pi^0 \pi^0} & =
   \frac{1}{\sqrt{2}} \left[ \lambda_u ( - C -
    P_{\tmop{tu}} + \hat{P}_{\tmop{EW}, \tmop{tu}} - \frac{2}{3} P_{\tmop{EW},
      \tmop{tu}}^C ) \right.
\nonumber\\
    &\left.- \lambda_c ( P - \hat{P}_{\tmop{EW}} + \frac{2}{3} P_{\tmop{EW}}^C
    ) \right]
\nonumber\\
  \bar{A}_{\pi^+ K^-}
  & =  \lambda_u^s ( T' - P'_{\tmop{tu}} - \frac{2}{3}
  P_{\tmop{EW}, \tmop{tu}}^{' C} ) - \lambda_c^s ( P' + \frac{2}{3}
  P_{\tmop{EW}}^{' C} )
  \nonumber\\
  \bar{A}_{\pi^0 \bar K^0}
  &=\frac{1}{\sqrt2}
  \left[
    \lambda^{s}_{u}(-C'-P'_{tu}+\hat{P}'_{EW,tu}-\frac{2}{3} P^{C}_{EW,tu})
        \right.
\nonumber\\
&\left. -\lambda^{s}_{c}(P'-\hat{P}'_{EW}+\frac{2}{3} P^{C}_{EW})
  \right]
\nonumber\\
  \bar{A}_{\pi^- \bar{K}^0}
  & =  \lambda_u^s ( P'_{\tmop{tu}} - \frac{1}{3}
  P^{' C}_{\tmop{EW}, \tmop{tu}} ) + \lambda_c^s ( P' - \frac{1}{3} P^{'
    C}_{\tmop{EW}} )
\nonumber\\
  \bar{A}_{\pi^0 K^-}
  &=-\frac{1}{\sqrt2}
  \left[
    \lambda^{s}_{u}(T'+C'-P'_{tu}-\hat{P}'_{EW,tu}+\frac{1}{3}P^{'C}_{EW})
        \right.
\nonumber\\
    & \left. -\lambda^{s}_{c}(P'+\hat{P}'_{EW}-\frac{1}{3}P^{'C}_{EW})
  \right]
\end{align}
where the rescaled amplitudes have a hierarchical structure $T \gg P \gg
\hat{P}_{\tmop{EW}}$. The primed and unprimed amplitudes are equal in the SU(3)
limit. For simplicity, throughout this paper, we will neglect the smallest
amplitudes of $P^{C}_{EW}$ and
take in a good approximation $P_{EW,tu}\simeq P_{EW,tc} = P_{EW}$ and
$P_{tu}\simeq P_{tc} = P$ due to t-quark dominance. As a phase convention, we
take $T$ to be real, i.e. $\delta_T = 0$. The amplitudes are normalized to the
CP averaged branching ratio $\tmop{Br} = ( |A|^2 + | \bar{A} |^2 )/2$ in units of $10^{-6}$, where the
tiny differences due to the $B^0$ and $B^{\pm}$ lifetime difference and the
final state phase spaces are neglected.
%Throughout the paper, all the $Br$s are
%written in unit of $10^{-6}$.
The direct CP violation is defined through
$a_{\tmop{CP}} = (| \bar{A} |^2 - |A|^2)/(|\bar A|^2 + |A|^2)$. The flavor SU(3)
symmetry breaking effects for amplitudes are considered as
$ |{T'}/{T}| = |{P'}/{P}| = |{P'_{EW}}/{P_{\tmop{EW}}}| \simeq
{f_K}/{f_{\pi}} \simeq 1.28 $ from naive factorization. The SU(3) symmetry breaking effects of strong
phases are characterized by the phase differences of the primed and the unprimed
amplitudes $ \Delta \delta_A \equiv \delta_A' - \delta_{A}$
with $A$ denoting for any of the amplitudes $T, P, P_{\tmop{EW}}$ etc.

\begin{figure}[htb]
%\begin{center}
\includegraphics[width=0.25\textwidth]{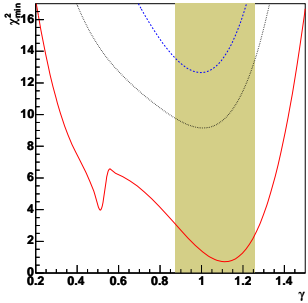}
%\end{center}
\caption{The $\chi_{min}^{2}$ as functions of $\gamma$. The three
curves (from bottom to top) are:
  Solid: Fit to $\pi^{+}\pi^{-}$ and $\pi^{+}K^-$ data only. Dashed: Fit to all
  the $\pi\pi$, $\pi K$ modes, with $\hat{P}_{EW}$ free (Fit B). Dotted: Fit to all the
  $\pi\pi$, $\pi K$ modes, with $\hat{P}_{EW}$ fixed by Eq.(\ref{Rew}) (Fit A). The shadowed
  band indicates the allowed range from the global SM fit.
}
\label{chiSq-A}
\end{figure}

%{\bf Weak phase $\gamma$ from $\pi^+ \pi^-$ and $ \pi^+ K^-$
%decays }.
The decay modes of $\pi^+ \pi^-$ and $\pi^+ K^-$ provide five data points: two
CP averaged branching ratios $Br(\pi^+ \pi^-) =4.6\pm0.4$ and $Br(\pi^+ K^-)
=18.2\pm0.9$, two direct CP asymmetries $a_{CP}(\pi^+\pi^-)$ and $a_{CP}(\pi^+
K^-)$, and one mixing induced CP asymmetry $S_{\pi\pi}=-0.61\pm 0.14$. Taking
the flavor $\tmop{SU} ( 3 )$ relations and neglecting $P^C_{EW}$, the two decay
modes only involves $T, P, \delta_P$ and the weak phase $\gamma$. Thus all of
them can be directly determined. A fit to the current data gives the following
results
\begin{eqnarray}\label{simpleFit}
 & &  |T| = 0.53\pm 0.03, \qquad |P| = 0.09 \pm 0.002,\nonumber \\
 & & \delta_P = - 0.48^{+ 0.09}_{- 0.12},  \qquad \gamma = 1.11^{+ 0.11}_{-
0.14}
\end{eqnarray}
with a $\chi^2_{\min}/d.o.f =0.71/1$. Where the well measured
result of $\sin 2\beta=0.73\pm 0.037$ from $B \rightarrow J / \psi K_S$ has been
used to relate the weak phase $\alpha$ to the weak phase $\gamma$ via
unitarity relation. The values of $|T|$ and $|P|$ are well determined with
relative errors less than $10 \%$.  The error of $\delta_P$ is larger but
can be reduced with more accurate data in the recent future. The ratio $|P / T|$ is
found to be around 0.17.  Note that the best fitted angle $\gamma$ is in a
remarkable agreement with the one from the global SM fit of the unitarity
triangle which gives $\gamma=1.08^{+0.17}_{-0.21}$ and at a compatible accuracy.
We emphasize that the above results are obtained without the interference with
other $\pi\pi$, $\pi K$ modes in which more diagrammatic parameters $C$ and
$P_{EW}$ are involved. Therefore it provides a very promising way to extract
$\gamma$ from charmless $B$ decays and an important reference point for any
further  analysis.
In obtaining the above result, the newly reported
$a_{CP}(\pi^+K^-)$ plays a key role. Without it, as shown in
ref\cite{SW,Buras}, the determination of $\gamma$ suffer from a
two-fold ambiguity with the other solution at $\gamma\simeq
40^\circ$. In Fig.\ref{chiSq-A}, we plot the $\chi^2_{min}$ as a
function of $\gamma$. It is clearly seen that after including
$a_{CP}(\pi^+K^-)$ the global minimum (best-fit) of $\chi^2$ falls
into the allowed range of the SM fit and the ambiguity is lifted.
%We plot in Fig.\ref{chiSq-A} the $\chi_{\min}^2$
%as functions of the weak phase $\gamma$.
%
%
%

%{\bf SU(3) symmetry  breaking effects of strong phase}.
%
Note that in the above fit the $\pi \pi$ and $\pi K$ modes are related via the
SU(3) relations, while the symmetry breaking effects on the strong phases have
been neglected. As pointed out in ref.  \cite{Zhou:2000hg} the
SU(3) breaking in strong phases may significantly change the correlation between
$a_{\tmop{CP}} ( \pi^+ \pi^- )$ and $a_{\tmop{CP}} ( \pi^+ K^- )$. Taking
$\Delta \delta_P$ as a free parameter in the fit, we find
\begin{align}\label{SF2}
 |T| &= 0.53\pm 0.03,  &|P|& = 0.09\pm 0.002
\nonumber \\
 \delta_P &= - 0.67^{+ 0.24}_{- 0.45},
 &\Delta \delta_p &= 0.21\pm0.4,
 \nonumber \\
\gamma &= 1.06 \pm 0.2 &&
\end{align}
%we yield a more consistent fit
%\begin{eqnarray}
%  &  & |T| = 0.56^{+ 0.040}_{- 0.038},  |P| = 0.093^{+ 0.0023}_{- 0.0025},
%   \delta_P = - 0.83^{+ 0.28}_{- 0.48}, \nonumber \\
%  & & \Delta \delta_p = 0.43^{+ 0.47}_{-0.31}, \gamma = 1.08 \pm 0.21
%\end{eqnarray}
%Here $\chi^2_{min}$  dramatically drops down to $\chi^2_{\min}
%=$0.19, which manifests  that a small value of $\Delta \delta_P$
%indeed improves considerably the goodness-of-fit.
with $\chi^2_{min}=4.7\times 10^{-7}$, which manifests  that a small 
value of $\Delta \delta_P$
further improves the goodness-of-fit.

%{\bf Large color-suppressed tree amplitude}.
When including the branching ratios of $Br(\pi^0\pi^0)=1.51\pm 0.28$ and
$Br(\pi^0\pi^-)=5.5\pm 0.6$ but ignoring $P_{\tmop{EW}}$ at the moment as both
modes are dominated by $T$ and $C$, only two new parameters $C$ and $\delta_C$
are involved. A fit to the four decay modes leads to the following results
\begin{eqnarray}\label{CFIT1}
& & |T| = 0.53^{+ 0.029}_{- 0.03},  |C| = 0.43\pm 0.05,
\delta_C = - 0.85^{+
0.52}_{-0.28} \nonumber \\
 & & |P| = 0.08^{+0.003}_{-0.005},  \delta_P = -0.48^{+ 0.09}_{- 0.11},
\gamma = 1.11^{+ 0.11}_{- 0.14}
\end{eqnarray}
which shows a large ratio of $|C / T| = 0.81$. In the QCD
factorization estimation this value is bound to be $|C / T| \le
0.4$. The error of $\delta_C$ is significantly large. Note that
the values of $|T|$ and $|P|$ and $\gamma$ remain almost
unchanged, which indicates  no explicit contradiction
between two sets of data $\pi^+ \pi^-, \pi^+ K^-$ and $\pi^0
\pi^0, \pi^0 \pi^-$, and the relatively large ratio $ |C / T|$ is
purely the results of the large $\pi^0\pi^0$ and $\pi^0\pi^-$
branching ratios.

%{\bf Extraction of electroweak penguin amplitude $\hat{P}_{\tmop{EW}}$}.
We now include other $\pi K$ data to determine $\hat{P}_{EW}$ and its strong
phase. We use the experimental value of $Br(\pi^0 \bar{K}^{0})=11.5\pm 1.0$,
$Br(\pi^-\bar{K}^0)=24.1\pm 1.3$ and $Br(\pi^{0}K^-)=12.1\pm 0.8$.  The
preliminary data of $a_{CP}(\pi^-\bar{K}^0)=0.02\pm 0.034$ and
$a_{CP}(\pi^{0}K^-)=0.04\pm 0.04$ are also considered. However, we do not
include the preliminary data of $a_{CP}(\pi^0\pi^0), a_{CP}(\pi^0\bar{K}^0)$ as
we would like to leave them to be pure predictions from the fits.
%as the current data are not yet conclusive
From the
isospin structure of the effective weak Hamiltonian in the SM and
the relations between the isospin amplitudes and the diagrammatic
amplitudes, i.e., $a^c_{2}= \hat{P}_{EW}$ and $a^u_{2}= T + C -
\hat{P}_{EW} $, one arrives at the following well-known
model-independent constraint \cite{EW}
\begin{eqnarray}
\frac{\hat{P}_{EW}}{T+C}\simeq  - \frac{3 ( C_9 + C_{10} )}{2 (
C_1 + C_2 )}\simeq ( 1.25\pm 0.12 ) \times 10^{- 2}
\label{Rew}
\end{eqnarray}
with $C_i s$ being the Wilson coefficients evaluated at $m_b$.
%In the SU(3) symmetry, it holds for primed amplitudes or the isospin
%$3/2$ part of $\pi K$ modes as well.
This relation tightly constrains the magnitude and the phase of
$\hat{P}_{\tmop{EW}}$. However in the presence of new physics
beyond the SM, the ratio could be significantly modified.
%When applying it to $\pi K$ modes, special care should
%be taken of the SU(3) breaking effects.
In view of the recent puzzling experimental results, a careful analysis is
urgently needed to find out whether this relation is indeed favored by the data.

% {\bf General global fitting}.
We now discuss several cases. First, consider a fit (Fit A) to the
$\pi\pi$, $\pi K$ data using Eq.(\ref{Rew}). The result is given
in the first column of Tab.\ref{charmlessFit}. Comparing with the
fit to $\pi^+\pi^-$ and $\pi^+ K^-$ in Eq.(\ref{simpleFit}), one
notices that the values of $\gamma$, $|T|$ and  $|P|$ are almost
unchanged. $C$ and its strong phase become larger and the ratio
$|C/T|$ is enhanced to be close to $\sim 0.9$. Namely, the large
$\pi^0\pi^0$ branching ratio is actually not the full reason for a
large $C/T\approx \mathcal{O}(1)$. It is also required by the $\pi
K$ data. The $\chi_{min}^{2}/d.o.f$ is found to be 12.7/7 which is
much higher than the previous fits. The main inconsistency comes
from the branching ratio of $\pi^+ K^-$, $\pi^0 K^0$ and $\pi^-
\bar{K}^0$. The resulting best-fit values in this case are
$20.0\pm0.8$, $9.7\pm0.5$ and $22.3\pm0.7$ respectively.
%which are much lower than the current data.
%
%%% new note added
The inconsistencies can be characterized by two ratios
$R_n=Br(\pi^+K^-)/Br(\pi^0\bar{K}^0) \simeq 0.79$ and
$R=Br(\pi^+K^-)/Br(\pi^-\bar{K}^0) \simeq 0.76$, which should be very close to
1.0 in the SM. The small value of $R_n$ may require corrections to $P_{EW}$ while
$R$ may be connected to large non-facotrizable effects \cite{Buras2}.
%%%%%
An important feature of this fit is that the predicted
direct CP violations $a_{CP}(\pi^{0}\pi^{0})\simeq 0.36$ and
$a_{CP}(\pi^0\bar{K}^0) \simeq -0.11$ are large and compatible with
$a_{CP}(\pi^{+}\pi^{-})$ and $a_{CP}(\pi^+\bar{K}^-)$. The
$\chi^2_{min}$ vs $\gamma$ is also given in Fig.\ref{chiSq-A}
which shows a good determination of $\gamma$.

\begin{table}[htb]%label{}
\begin{center}
\begin{ruledtabular}
\begin{tabular}{lllll}
  & FitA & FitB & FitC & FitD\\
  \hline
  $\gamma$                 & $1.0^{+ 0.11}_{- 0.13}$   & $1.0^{+ 0.13}_{- 0.18}$   & $0.98^{+0.12}_{0.13}$           & $1.1^{+ 0.12}_{- 0.19}$\\
  \hline
  $|T|$                    & $0.52 \pm 0.27$           & $0.52 \pm 0.03$           & $0.52 \pm 0.03$           & $1.13^{+ 0.36}_{- 0.32}$\\
  \hline
  $|C|$                    & $0.47 \pm 0.04$           & $0.45 \pm 0.05$           & $0.45\pm 0.05$   & $0.32^{+0.35}_{-0.22}$\\
  \hline
  $\delta_C$               & $-1.1^{+ 0.19}_{- 0.17}$  & $-0.88^{+ 0.3}_{- 0.2}$   & $-1.87^{+0.3}_{-0.25}$     & $-2.7^{+1.29}_{-0.3}$\\
  \hline
  $|P|$                    & $0.094 \pm 0.001$         & $0.093 \pm 0.002$         & $0.09 \pm 0.002$          & $0.74\pm 0.3$\\
  \hline
  $\delta_P$               & $-0.49^{+ 0.09}_{- 0.10}$ & $-0.53^{+ 0.10}_{- 0.14}$ & $-0.76 \pm 0.17$ & -$0.2^{+ 0.05}_{- 0.14}$\\
  \hline
  $|\hat{P}_{\tmop{EW}} |$ & $-$                       & $0.03 \pm 0.01$           & $0.03 \pm 0.01$           & $0.024\pm 0.01$\\
  \hline
  $\delta_{P_{\tmop{EW}}}$ & $-$                       & $0.67^{+ 0.2}_{- 0.3}$    & $0.67^{+0.2}_{-0.4}$    & $1.13^{+0.19}_{-0.39}$\\
  \hline
  $|P_D|$                  & $0$(fix)                  & $0$(fix)                  & $0$(fix)                  & $0.11\pm 0.02$\\
  \hline
  $\delta_{P_D}$           & $0$(fix)                  & $0$(fix)                  & $0$(fix)                  & $- 0.21^{+ 0.09}_{-0.14}$\\
  \hline
  $\Delta \delta_P$        & $0$(fix)                  & $0$(fix)                  & $0.2^{+ 0.1}_{- 0.17}$    & $0$(fix)\\
  \hline
  $\chi^2/\mbox{dof}$    & $12.7/7$                  & $9.1/5$                   & 7.9/4                     & 5.4/3\\
  \hline
  \hline
  $a_{\pi^0 \pi^0}$        & $0.36 \pm 0.11$           & $0.05 \pm 0.22$           & -$0.06 \pm 0.2$          & $0.07 \pm 0.39$\\
  \hline
  $a_{\pi^0 \bar{K}^0}$    & -$0.10 \pm 0.004$        & -$0.01\pm 0.05$          & -$0.02 \pm 0.05$         & -$0.01 \pm 0.11$\\
  \hline
  \hline
  $B_{\pi^0 \pi^0}$        & $1.7 \pm 0.3$             & $1.56 \pm 0.4$            & $1.53 \pm 0.4$           & $1.7\pm0.5$\\
  \hline
  $B_{\pi^0 \bar{K}^0}$     & $9.7 \pm 0.48$            & $11.1 \pm 1.8$            & $11.1 \pm 2.1$            & $11.3\pm2.3$\\
  \hline
  $B_{\pi^0 K^-}$          & $11.7 \pm 0.6$            & $11.9 \pm 2.2$            & $11.8 \pm 2.4$            & $11.9\pm2.5$\\
  \hline
  $a_{\pi^+ \pi^-}$        & $0.27 \pm 0.06$           & $0.30 \pm 0.08$            & $0.37 \pm 0.06 $             & $0.34\pm0.27$\\
  \hline
  $a_{\pi^+ K^- }$         & $-0.1 \pm 0.02$           & $-0.11 \pm 0.02$          & $-0.1 \pm 0.03$          & -$0.1\pm 0.06$\\
%  \hline
\end{tabular}
\end{ruledtabular}
\end{center}
\caption{Best fitted parameters and predictions from  charmless
$B$ decay data in four different cases.
Details are explained in the text.}
\label{charmlessFit}
\end{table}
Second, considering a fit (Fit B) with freeing the parameter
$\hat{P}_{\tmop{EW}}$ and its strong phase. The results are
tabulated in the second column of table \ref{charmlessFit}, which
show roughly the same values of $\gamma$, $|T|$, $|P|$ and $|C|$,
but the value of $|P_{\tmop{EW}}|\simeq 0.03$ leads to
\begin{equation}
\frac{|\hat{P}_{\tmop{EW}}|}{ |T+C|} \simeq (3.1\pm 1.3)\times 10^{-2}
\end{equation}
which is twice as large as in Eq.(\ref{Rew}).
%In this case,
%a much smaller $\chi_{\min}^2 =2.9$ indicates an improved
%consistency. Consequently,
The  data of $\tmop{Br} ( \pi^0 \bar{K}^0 )$ and the ratio $R_n$
are perfectly reproduced. This result agrees with the observation
in Refs.\cite{Buras} with a statement that a large electro-weak
penguin can consistently explain the $\pi K$ data. Clearly, the
large value $\hat{P}_{EW}$ is driven by the observed large
branching ratio of $\pi^{0} \bar{K}^{0}$ mode.  All the previous
fits with small $\hat{P}_{EW}$ failed to  meet this date
point\cite{Rosner,Hocker}. Note that in the case of large
$\hat{P}_{EW}$, the predicted CP violations of $\pi^{0}\pi^{0}$
and $\pi^{0}\bar{K}^0$ are found to be small. The predicted
central values are only $0.06$ and $-0.02$ respectively though the
errors are big(see Tab.\ref{charmlessFit}). Therefore it provides
a possibility to distinguish the electro-weak penguin effects in
the near future with more accurate measurements.
%To see how
%
In the third column (Fit C) of Tab.\ref{charmlessFit}, we
consider the effects of $\tmop{SU} ( 3 )$ breaking in the strong
phases. The best fit gives $\Delta \delta_P = 0.2$ in accordance
with Eq.(\ref{SF2}).
%which
%coincides with the one obtained from the $\pi^+ \pi^-$ and $\pi^+
%K^-$ modes only.
In this case,
%all the experimental data are perfectly reproduced and the
value of $\hat{P}_{EW}$ remains the same as in Fit.B.
%But a smaller $\gamma$ is favored.
The predictions give $a_{CP}(\pi^0\pi^0)=-0.06$ and
$a_{CP}(\pi^0\bar{K}^0)=-0.1$ respectively.
%

%
%{\bf Inelastic rescattering effects}.
The inclusion of all the
$\pi \pi$ and $\pi K$ modes allows one to investigate the possible
large inelastic rescattering effect due to the process of $B
\rightarrow D D_{( s )} \rightarrow \pi \pi ( \pi K )$.  It is
well known that $B \rightarrow D D$ have a large branching ratio
about 50 times greater than that of $B \rightarrow \pi \pi$, which
amplifies the successive small effects of re-scattering $D D_{( s
)} \rightarrow \pi \pi ( \pi K )$.
%This effect was discussed
%recently in Refs. \cite{DD} in  model dependent and
%independent ways.
%
Considering the fact that $B \rightarrow D D_{( s )}$ only
contributes to the isospin $0 ( 1 / 2 )$ $\pi\pi(\pi K)$ final
sates and carries only the CKM factor $\lambda^{(s)}_c$,
its contribution can be parameterized by only one complex quantity
denoted by $D(D')$ and effectively it can be considered by
replacing $P^{(')}$ by $P^{(')}_D = P^{(')} + D^{(')}$.
In the fourth  column (Fit D) of Tab.\ref{charmlessFit}, the
parameters of $|P_{D}|$ and $\delta_{P_{D}}$ motivated by the inelastic
rescattering from $B \rightarrow D D_{( s )}$ are added
which makes $P$ and $P_{D}$ two independent parameters.  The results show that
$P_{D}$ is compatible with the QCD penguins obtained from the previous fits of
A,B,C while $P$ becomes larger.  This large difference between $P$ and $P_{D}$
indicates a large effect of inelastic rescattering and may also imply new
physics in strong penguin sector.  In this fit, the ratio of $|C/T|$ is reduced
to $0.37$. The ratio of $\hat{P}_{EW}/(T+C)$ remains large and the two predicted CP
violations are again small.

In conclusion, the current data enable us to make a very encouraging global
fitting for testing the standard model and probing new physics.
%It will be very
%helpful and crucial if the Belle and BaBar collaborations can arrive at more
%accurate measurements on both branching ratios and direct CP violations in $B\to
%\pi^0\pi^0$ and $\pi^0 \bar{K}^0$ decays.
It will be very crucial to arrive at more accurate measurements on
both branching ratios and direct CP violations in $B\to
\pi^0\pi^0$ and $\pi^0 \bar{K}^0$. The current preliminary data
give $a_{CP}(\pi^0\pi^0)=0.28\pm0.39$ and $a_{CP}(\pi^0
\bar{K}^0)=-0.09\pm0.14$\cite{newData}. Due to the large errors,
including them will not change the conclusion. Numerically, we
find that the results in Eq.(\ref{CFIT1}) are unchanged.  The
ratio $|\hat{P}_{EW}/(T+C)|$ remains large and is found to be
$0.024\pm0.01$, $0.034\pm0.01$ and $0.033\pm0.04$ for FitB,C and D
respectively in Tab.\ref{charmlessFit}.
It is very likely that we are standing
at the corner of finding new physics with two B-factories.

acknowledgments: YLW is supported in part by the key project of
NSFC and Chinese Academy of Sciences.
%%%%%%%%%%%%%%%%%%%%%%%%%%%%%%%%%%%%% reference %%%%%%%%%%%%%%

%%%%%%%%%%%%%%%%%%%%%%%%%%%%%%%%%%%%%%%%%%%%%%%%%%%%%

\begin{thebibliography}{50}


\bibitem{newData}
K.Abe. ~et.al (Belle) Phys.Rev.{\bf D}012001(2003).
B.Aubert, ~et.al(Babar), hep-ex/0407057.
Z.Ligeti, ~talk at ICHEP04,Beijing,Auguest 16-22(2004),hep-ph/0408267.
M. Giorgi, ~talk at ICHEP04, hep-ex/0408113,
Y. Sakai,~talk at ICHEP04, hep-ex/0410006.
\bibitem{Wu:2000ki}
Y.~L.~Wu,
%``A new prediction for direct CP violation epsilon'/epsilon and Delta(I)  = 1/2
%rule,''
Phys.\ Rev.\ D {\bf 64}, 016001 (2001)
%[arXiv:hep-ph/0012371].
%%CITATION = HEP-PH 0012371;%%

\bibitem{HFaverage}
For a summary, see: Heavy Flavor Averaging Group,
http://www.slac.stanford.edu/xorg/hfag/rare

\bibitem{S2HDM}
Y.L. Wu and L. Wolfenstein, Phys.Rev.Lett. {\bf 73} 1762 (1994);
%L. Wolfenstein and Y. L. Wu,
Phys.Rev.Lett. {\bf 73} 2809 (1994);
Y.L. Wu and Y.F. Zhou, Phys.Rev. {\bf D61} 096001 (2000).

\bibitem{SW} J. Silva and L. Wolfenstein, Phys.Rev. {\bf D49} 1151
(1994)

\bibitem{Buras}
A.~J.~Buras {\it et al.},
%``B $\to$ pi pi, new physics in B $\to$ pi K and implications for rare K and B
%decays,''
Phys.\ Rev.\ Lett.\  {\bf 92}, 101804 (2004).
%%CITATION = HEP-PH 0312259;%%
hep-ph/0402112.
%%CITATION = HEP-PH 0402112;%%
T.~Yoshikawa,
%``A possibility of large electro-weak penguin contribution in B $\to$ K pi
%modes,''
Phys.\ Rev.\ D {\bf 68}, 054023 (2003)
%[arXiv:hep-ph/0306147].
%%CITATION = HEP-PH 0306147;%%
D.~Atwood and G.~Hiller,
%``Implications of non-standard CP violation in hadronic B decays,''
hep-ph/0307251.
%%CITATION = HEP-PH 0307251;%%
S.~Mishima and T.~Yoshikawa,
%``Large electroweak penguin contribution in B $\to$ K pi and pi pi decay
%modes,''
hep-ph/0408090.
%%CITATION = HEP-PH 0408090;%%
S.~Nandi and A.~Kundu,
%``Large electroweak penguins in B $\to$ pi pi and B $\to$ pi K: Implication
%for new physics,''
hep-ph/0407061.
%%CITATION = HEP-PH 0407061;%%
D. Atwood and A. Soni, Phys. Rev.{\bf D58}, 036005 (1998). 

\bibitem{Rosner}
C.~W.~Chiang {\it et al.},
%``Charmless B $\to$ P P decays using flavor SU(3) symmetry,''
Phys.\ Rev.\ D {\bf 70}, 034020 (2004).
%[arXiv:hep-ph/0404073].
%%CITATION = HEP-PH 0404073;%%
Phys.\ Rev.\ D {\bf 69}, 034001 (2004).
%
%

\bibitem{Hocker}
J.~Charles {\it et al.},  %[CKMfitter Group Collaboration],
%``CP violation and the CKM matrix: Assessing the impact of the asymmetric B
%factories,''
arXiv:hep-ph/0406184.
%%CITATION = HEP-PH 0406184;%%

\bibitem{Zhou:2000hg}
Y.~L.~Wu and Y.~F.~Zhou,
%``Weak phase gamma and strong phase delta from CP averaged B $\to$ pi pi  and
%pi K decays,''
Phys.\ Rev.\ D {\bf 62}, 036007 (2000).
%[arXiv:hep-ph/0002227].
%%CITATION = HEP-PH 0002227;%%
%
Eur.\ Phys.\ J.\ dirC {\bf 5}, 014 (2003).
%%CITATION = HEP-PH 0210367;%%
%
Y.~F.~Zhou, et.al,%Y.~L.~Wu, J.~N.~Ng and C.~Q.~Geng,
%``The interplay of weak and strong phases and direct CP violation in  charmless
%B meson decays,''
Phys.\ Rev.\ D {\bf 63}, 054011 (2001).
%[arXiv:hep-ph/0006225].
%%CITATION = HEP-PH 0006225;%%
%

\bibitem{EW}
M. Gronau, et.al, %Dan Pirjol, Tung-Mow Yan,.
Phys.Rev.D60:034021(1999).
%
Y.Grossman, M.Neubert and  A.L.Kagan, JHEP 9910:029(1999)

\bibitem{He}
X.~G.~He, et.al, %Y.~K.~Hsiao, J.~Q.~Shi, Y.~L.~Wu and Y.~F.~Zhou,
%``The CP violating phase gamma from global fit of rare charmless hadronic  B
%decays,''
Phys.\ Rev.\ D {\bf 64}, 034002 (2001)
%[arXiv:hep-ph/0011337].
%%CITATION = HEP-PH 0011337;%%

\bibitem{QCDFacFit}
M.~Beneke and M.~Neubert,
%``QCD factorization for B $\to$ P P and B $\to$ P V decays,''
Nucl.\ Phys.\ B {\bf 675}, 333 (2003)
%[arXiv:hep-ph/0308039].
%%CITATION = HEP-PH 0308039;%%
%
D.~s.~Du, et.al, %J.~f.~Sun, D.~s.~Yang and G.~h.~Zhu,
%``Charmless two-body B decays: A global analysis with QCD factorization,''
Phys.\ Rev.\ D {\bf 67}, 014023 (2003)
%[arXiv:hep-ph/0209233].
%%CITATION = HEP-PH 0209233;%%
%
R.~Aleksan, et.al,%P.~F.~Giraud, V.~Morenas, O.~Pene and A.~S.~Safir,
%``Testing QCD factorisation and charming penguins in charmless B $\to$ P V.
%((U)),''
Phys.\ Rev.\ D {\bf 67}, 094019 (2003).
%[arXiv:hep-ph/0301165].
%%CITATION = HEP-PH 0301165;%%
G.~Buchalla and A.~S.~Safir,
%``Model independent bound on the unitarity triangle from CP violation in B
%$\to$ pi+ pi- and B $\to$ psi K(S),''
Phys.\ Rev.\ Lett.\  {\bf 93}, 021801 (2004).

\bibitem{pQCD}
Y.~Y.~Keum, H.~n.~Li and A.~I.~Sanda,
%``Fat penguins and imaginary penguins in perturbative QCD,''
Phys.\ Lett.\ B {\bf 504}, 6 (2001)
%[arXiv:hep-ph/0004004].
%%CITATION = HEP-PH 0004004;%%
%
%Y.~Y.~Keum, H.~N.~Li and A.~I.~Sanda,
%``Penguin enhancement and B $\to$ K pi decays in perturbative QCD,''
Phys.\ Rev.\ D {\bf 63}, 054008 (2001)
%[arXiv:hep-ph/0004173].
%%CITATION = HEP-PH 0004173;%%

\bibitem{SCET} C.W. Bauer et. al.,  hep-ph/0401188.

\bibitem{Buras2}
A.~J.~Buras {\it et al.}, hep-ph/0410407.

%\bibitem{DD}
%see, e.g.
%A.~N.~Kamal and C.~W.~Luo,
%%``Inelastic final-state interactions and two-body hadronic B decays into
%%single-isospin channels,''
%Phys.\ Rev.\ D {\bf 57}, 4275 (1998).
%%[arXiv:hep-ph/9710275].
%%%CITATION = HEP-PH 9710275;%%
%S.~Barshay, G.~Kreyerhoff and L.~M.~Sehgal,
%%``Direct CP violation in B-+ $\to$ pi-+ omega, pi-+ rho0, pi0 rho-+, and in
%%anti-B0 (B0) $\to$ pi-+ rho+- with an enhanced branching ratio for pi0 rho0,''
%Phys.\ Lett.\ B {\bf 595}, 318 (2004).
%%[arXiv:hep-ph/0405012].
%%%CITATION = HEP-PH 0405012;%%
\end{thebibliography}
\end{document}